\documentclass[aps,prd,reprint,superscriptaddress]{revtex4-2}

\usepackage{graphicx}
\usepackage{float}
\makeatletter
\let\newfloat\newfloat@ltx
\makeatother
\usepackage{subfig}
\usepackage{dcolumn}
\usepackage{lmodern}
\usepackage{bm}

\usepackage[utf8]{inputenc}
\usepackage[T1]{fontenc}
\usepackage{booktabs, array, mathptmx, tabularx, booktabs, lipsum, amsmath,multirow}
\usepackage{siunitx, xcolor}
\usepackage[version=4]{mhchem}

\usepackage{braket}
\usepackage{algorithm}
\usepackage{algorithmicx}
\usepackage[noend]{algpseudocode}
\usepackage{microtype}
\usepackage[colorlinks,linkcolor=blue,anchorcolor=blue,citecolor=blue]{hyperref}

\begin{document}


\title{Quantum Search for Gravitational Wave of Massive Black Hole Binaries}


\author{Fangzhou Guo}
\email[]{guofangzhou23@mails.ucas.ac.cn}
\affiliation{School of Fundamental Physics and Mathematical Sciences, Hangzhou Institute for Advanced Study, UCAS, Hangzhou 310024, China}
\affiliation{University of Chinese Academy of Sciences, Beijing 100049, China}
\affiliation{Institute of Theoretical Physics, Chinese Academy of Sciences, Beijing 100190, China}
\affiliation{International Centre for Theoretical Physics Asia-Pacific, UCAS, Beijing 100190, China}
\author{Jibo He}
\email[]{jibo.he@ucas.ac.cn}
\affiliation{School of Fundamental Physics and Mathematical Sciences, Hangzhou Institute for Advanced Study, UCAS, Hangzhou 310024, China}
\affiliation{University of Chinese Academy of Sciences, Beijing 100049, China}
\affiliation{International Centre for Theoretical Physics Asia-Pacific, UCAS, Beijing 100190, China}
\affiliation{Taiji Laboratory for Gravitational Wave Universe (Beijing/Hangzhou), UCAS, Beijing 100190, China}

\date{\today}

\begin{abstract}
Matched filtering is a common method for detecting gravitational waves. However, the computational costs of searching large template banks limit the efficiency of classical algorithms when searching for massive black hole binary (MBHB) systems. This work explores the application of a quantum matched filtering algorithm based on Grover’s algorithm to MBHB signals. Under certain simplifying assumptions, quantum approach can reduce the computational complexity from $O(N)$ to $O(\sqrt{N})$ theoretically, where $N$ is the size of the template bank. Simulated results illustrate the potential reduction in computational costs, while also showing that the performance can degrade in some cases due to instability of the algorithm. These findings reveal both the potential and the limitations of quantum search for MBHB signals, pointing to the importance of developing more robust and stable search strategies alongside realistic modeling in future work.
\end{abstract}


\maketitle

\section{\label{sec:level1}Introduction}

Gravitational waves (GWs), predicted by Einstein’s theory of general relativity, have become a powerful tool for exploring astrophysical phenomena. Since the first detection of GWs from a binary black hole merger by LIGO in 2015~\cite{abbott2016observation}, this field has advanced rapidly. However, the ground-based detectors like LIGO are limited to high-frequency gravitational waves (10~Hz to kHz), leaving a significant portion of the gravitational wave spectrum, particularly the low frequency band inaccessible. To address this limitation, space-based detectors such as the Laser Interferometer Space Antenna (LISA)~\cite{amaro2017laser}, Taiji~\cite{hu2017taiji}, and TianQin~\cite{luo2016tianqin} are currently under development, opening an entirely new observational window on the Universe by around 2030. One of the most promising future sources of GWs is the coalescence of massive black hole binaries (MBHBs), which are expected to be detected by the upcoming space-based detectors. MBHB mergers, occurring at the centers of galaxies, produce strong gravitational wave signals that carry valuable information about black hole masses, spins, and the dynamics of galaxy formation and evolution~\cite{sesana2013gravitational}.  

Nowadays, matched filtering technique is commonly used for detecting GW signals in ground-based GW observations~\cite{owen1999matched}. This approach has been developed and applied in several GW search pipelines, such as the GstLAL pipeline~\cite{ewing2024performance} and the PyCBC pipeline~\cite{usman2016pycbc}, and hierarchical search strategies~\cite{gadre2019hierarchical} have been proposed to improve computational efficiency. Matched filtering also provides a foundation for parameter estimation~\cite{poisson1995gravitational,christensen2022parameter}. Comprehensive reviews of GW data analysis and matched filtering methods can be found in Ref.~\cite{jaranowski2012gravitational}.

However, detecting MBHB signals using matched filtering is a significant computational challenge. MBHB signals are characterized by a high-dimensional parameter space, which typically spans around 15 dimensions, depending on the inclusion of spin and orbital effects~\cite{lang2008localizing}, which makes the template bank of MBHB signals significantly large. Early estimates suggested that the template bank size could reach $10^{13}$ for non-spinning MBHB searches, making classical matched filtering computationally prohibitive~\cite{cornish2007search, cornish2007catching}. However, classical methodological advances have since improved its feasibility, making it a crucial tool in milli-Hz GW parameter estimation~\cite{katz2020gpu,hoy2024rapid}.

Quantum computing has also emerged as a promising method for solving this computational bottleneck. Grover’s algorithm offers a quadratic speedup for searching unsorted databases, reducing the complexity from $O(N)$ to $O(\sqrt{N})$~\cite{grover1996fast}. Previous study have demonstrated that Grover's algorithm can be applied to matched filtering to improve gravitational wave signal detection and the efficiency of the quantum matched filtering algorithm has been tested on the GW150914 event~\cite{gao2022quantum}. But its application to MBHB signals remains unexplored. Given the large template bank size and the complex waveform of MBHB signals, applying Grover's algorithm to MBHB matched filtering has the potential to enhance computational efficiency and improve detection capabilities.

In this work, we explore the application of quantum matched filtering approach for MBHB signal detection, suggesting that, in principle, Grover's algorithm could offer a more efficient way to search large MBHB template banks. Through numerical simulations, the detection performance and computational costs of the quantum-enhanced method are evaluated.

This paper is organized as follows: Section \ref{sec:2} provides a background on classical matched filtering and Grover's algorithm. Section \ref{sec:3} details the quantum algorithm and its implementation. Section \ref{sec:4} gives a brief discussion on the result and suggesting further directions for study.

\section{Methodology} \label{sec:2}
The algorithm presented in this work builds upon matched filtering, Grover's algorithm and quantum counting, following the detailed introduction and application of these methods in Ref.~\cite{gao2022quantum}. For readers interested in the underlying quantum computing concepts, comprehensive explanations of Grover's algorithm and quantum counting can be found in textbooks as Ref.~\cite{mcmahon2008quantum,nielsen2010quantum,yanofsky2008quantum}. In addition, the theory and application of matched filtering are well established in classical signal processing literature~\cite{helstrom1960statistical,van2004detection}. A brief overview of these methods is provided below for clarity.

\subsection{Classical Matched Filtering}
Matched filtering is a signal processing technique designed to optimally detect known signals buried in additive, stationary Gaussian noise~\cite{helstrom1960statistical}. Given a known signal template $h(t)$ and detector data $s(t) = h(t) + n(t)$, in which $n(t)$ is the additive noise, matched filtering maximizes the signal-to-noise ratio (SNR) by correlating the data with a time-reversed and noise-weighted version template~\cite{allen2005chi}.

In the frequency domain, the matched filter output is computed as
\begin{equation}
\tilde{z}(f) = \frac{\tilde{s}(f)\tilde{h}^*(f)}{S_n(f)},
\end{equation}
where $\tilde{s}(f)$ and $\tilde{h}(f)$ are the Fourier transforms of the data and the template, respectively, and $S_n(f)$ is the one-sided power spectral density of the detector noise. 

To cover a broad range of potential signal parameters, a discrete set of templates $\{h_i(t)\}$ is constructed, known as a \textit{template bank}. For each template, the detection statistic $\rho_i(t)$ is computed and scanned for peaks. The detection statistic is typically defined as maximum SNR:
\begin{equation}
\rho_{i}(t) =  \frac{\langle s | h_i \rangle}{\sqrt{\langle h_i | h_i \rangle}},
\end{equation}
where the inner product is defined as
\begin{equation}
\langle a | b \rangle = 4\,\mathrm{Re} \int_0^\infty \frac{\tilde{a}(f)\tilde{b}^*(f)}{S_n(f)} df.
\end{equation}

If the maximum value of $\rho_i(t)$ exceeds a preset threshold $\rho_{\text{thr}}$, the template is considered to match the signal, indicating a potential detection. The statistic $\rho_{i}(t)$ can be efficiently computed using the Fast Fourier Transform (FFT) algorithm with computational complexity $O(M\log M)$~\cite{cooley1967historical}, where $M$ is the number of time samples.
The overall computational cost of applying matched filtering over a template bank of $N$ templates is thus $O(N M \log M)$.

\subsection{Grover’s Algorithm} \label{grover}
Here is a short geometric illustration, which is sufficient to understand the effectiveness of Grover's algorithm. 

Grover's algorithm provides a quantum solution to the unstructured search problem. Given a search space of size $N$ and $r$ marked solutions such that $f(x) = 1$ (for non-solutions, $f(x)=0$), the goal is to find one such $x$ efficiently. Classically, the complexity of this problem is $O(N)$~\cite{barnett2009quantum}, while Grover's algorithm offers a quadratic speedup with a complexity of $O(\sqrt{N})$~\cite{grover1996fast}.

The algorithm begins by preparing a uniform superposition over all basis states:
\begin{equation}
\ket{s} = \frac{1}{\sqrt{N}} \sum_{i=0}^{N-1} \ket{i}.
\end{equation}
The first key component of Grover's algorithm is the \textit{oracle} operator $U_f$, which encodes the function $f(x)$ by flipping the sign of the marked states:
\begin{equation} \label{kick}
U_f \ket{x} = (-1)^{f(x)} \ket{x}.
\end{equation}
This means that if $x$ is a solution, the amplitude of $\ket{x}$ is inverted in phase, while all other basis states remain unchanged.

Geometrically, Grover’s algorithm performs a rotation in the two-dimensional subspace spanned by $\ket{\alpha}$ (the equal superposition of unmarked states) and $\ket{\beta}$ (the equal superposition of marked states). The initial state $\ket{s}$ lies at an angle $\theta$ from $\ket{\alpha}$, where
\begin{equation}\label{theta_r}
\sin(\theta) = \sqrt{\frac{r}{N}}.
\end{equation}

Each Grover iteration consists of applying the oracle $U_f$ followed by the Grover diffusion operator $D = 2\ket{s}\bra{s} - I$. Together, they form the Grover operator
\begin{equation}
G = D \cdot U_f,
\end{equation}
which rotates the state vector by an angle of $2\theta$ toward $\ket{\beta}$.
After approximately
\begin{equation} \label{optk_appro}
k \approx \left\lfloor \frac{\pi/2-\theta}{2\theta}  \right\rfloor \approx \left\lfloor \frac{\pi}{4} \sqrt{\frac{N}{r}} - \frac{1}{2}\right\rfloor
\end{equation}
iterations, the state vector is close to $\ket{\beta}$, and a measurement yields a correct result with high probability, completing the quantum search in $O(\sqrt{N})$ steps.

\subsection{Quantum Counting} \label{subsec:QC}
In Grover's algorithm, the number of iterations $k$ depends on the number of solutions $r$, which is often unknown in practical problems such as gravitational wave matched filtering. To address this, one can estimate $r$ using quantum counting. Since $r$ can be obtained from $\theta$ via Eq.~\eqref{theta_r}, the key is to estimate $\theta$ using quantum phase estimation (QPE)~\cite{kitaev1997quantum}, an application of quantum Fourier transform (QFT)~\cite{kaye2006introduction}.

The Grover operator $G$ has eigenvalues $e^{\pm i2\theta}$ in the subspace spanned by $\ket{\alpha}$ and $\ket{\beta}$.
The quantum counting procedure applies QPE to $G$ using an ancillary register of $p$ qubits to estimate $\theta$. This involves initializing the ancilla in a uniform superposition, applying controlled-$G^{2^j}$ operations, and performing the inverse quantum Fourier transform. Measuring the ancilla yields an estimate of $\theta$. The most probable outcome $\ket{l'}$ satisfies
\begin{align} \label{sdfg}
	\frac{\theta}{\pi} \approx \frac{l'}{2^p}, \quad \text{or} \quad \frac{\pi - \theta}{\pi} \approx \frac{l'}{2^p}.
\end{align}
Typically, when $r/N \ll 1$, the smaller angle corresponds to the correct estimate. Letting $b$ be the measured outcome, the estimated angle is
\begin{align}
	\theta_* = 
	\begin{cases}
		\frac{b \pi}{2^p}, & b \leq 2^{p-1}, \\
		\pi - \frac{b \pi}{2^p}, & b > 2^{p-1}.
	\end{cases}
\end{align}

The estimation error depends on the precision of the phase estimation. For a desired precision of $m$ bits and error probability $\epsilon$, the required number of qubits is~\cite{nielsen2010quantum}:
\begin{align}
	p = m + \log \left(2 + \frac{1}{2\epsilon} \right).
\end{align}
If the target precision for $\theta$ is $O(1/\sqrt{N})$, then $m = O(\frac{1}{2} \log N)$ and hence $p = O(\log N)$. The number of applications of $G$ is $2^p - 1 = O(\sqrt{N})$, so the total complexity remains $O(\sqrt{N})$. Once $\theta$ is estimated, we can compute $r$ and determine the optimal iteration number $k$ for Grover's algorithm.

\section{Quantum Matched Filtering Applied to MBHB} \label{sec:3}
A brief introduction to the quantum matched filtering method 
as proposed in Ref.~\cite{gao2022quantum} is now given, followed by a discussion on how this method can be applied in the context of MBHBs.

In Ref.~\cite{gao2022quantum}, two quantum algorithms are proposed for gravitational wave matched filtering. The first uses quantum counting to check whether any matching signal exists—addressing a core question in gravitational wave detection. The second applies Grover’s algorithm to identify a matching template, when at least one match is present.

For convenience, the same notation as in Ref.~\cite{gao2022quantum} is used: the number of templates is denoted by $N$, and the number of data points in the time-series by $M$. They chose a digital encoding to represent the data and templates as classical bits encoded in the computational basis. They outlined a specific implementation, making use of four registers: one data register $\ket{D}$ which must be of size linear in $M$, and one index register $\ket{I}$, which requires $\log N$ qubits. For intermediate calculations they specify also one register to hold the computed template $\ket{T}$, which must be of size linear in $M$, and one $\ket{\rho}$to hold the computed SNR value, which does not scale with $N$ or $M$ and is $O(1)$. Although the SNR register is denoted as $O(1)$ in the resource estimates, in practice it must be implemented with finite precision. This requires a fixed number of qubits, $n_\text{SNR}$, to encode the matched-filter output when using a fixed-point representation. Insufficient bit-width may lead to rounding or comparator errors, which can in turn cause deviations in the acceptance probability of true signals (false negatives) or spurious detections (false positives). The required precision depends on the desired detection fidelity; importantly, it does not scale with the template bank size $N$ or the data length $M$. Therefore, while finite precision introduces an additional overhead to the oracle, it does not alter the asymptotic scaling advantage discussed in this work.

A key aspect of oracle construction, according to Ref.~\cite{gao2022quantum}, is that gravitational-wave signals are accurately described by general relativity, allowing the templates to be computed on-the-fly during the matched filtering process. This eliminates the need to pre-load templates into quantum random access memory (qRAM)~\cite{giovannetti2008architectures} and thus avoids the hidden complexity associated with this loading step, as well as concerns about the experimental feasibility of constructing qRAM~\cite{aaronson2015read,preskill2018quantum}. The steps for constructing an oracle that checks template matches are already part of classical analysis; explicitly including them does not reduce the potential quantum speed-up.

The cost of a single oracle call (one SNR calculation) is non-negligible and must be included in a full complexity analysis. While Grover’s algorithm does not accelerate this step but improves the overall scaling with the number of templates.

For clarity, we note that the classical algorithm can also be described in terms of an oracle. The classical oracle must first compute the template waveform from the index, which takes time linear in the number of data points $M$. It then calculates the SNR between the template and the data, which requires $O(M \log M)$ operations, and finally checks whether the result exceeds a given threshold, which takes $O(1)$ gates. To compute the match for all $N$ templates, the total number of calls to the classical oracle is proportional to $N$, giving a classical computational complexity of $O(N M \log M)$.

\begin{algorithm}[htbp]
	\caption{Grover's Gate \newline Complexity: $O(M\log M+\log N)$~\cite{gao2022quantum}}\label{alg:GroGate}
	\begin{algorithmic}[1]
		\Function{Grover's Search algorithm}{$N$, $\ket{D}$, $\rho_{\textrm{thr}}$}
		\label{GroverPseudo}
		\Procedure{Oracle Construction}{}
		\State \emph{Creating templates}:\label{alg:step1S}
		\ForAll{$i<N$}
		\State{$ \ket{i}\ket{0} \gets \ket{i}\ket{T_i}$}
		\EndFor\label{alg:step1E}
		\State \emph{Comparison with the data}:\label{alg:step2S}
		\State$\ket{i}\ket{D}\ket{T_i}\ket{0} \gets \ket{i}\ket{D}\ket{T_i}\ket{\rho(i)}$
		\If {$\rho(i) <  \rho_{\textrm{thr}}$}
		\State $f(i)=0$
		\Else 
		\State $f(i)=1$
		\EndIf
		$\ket{i}\ket{D}\ket{T_i}\ket{\rho(i)} \gets (-1)^{f(i)}\ket{i}\ket{D}\ket{T_i}\ket{\rho(i)} $ \label{alg:step2E}
		\State \emph{Dis-entangling registers}:\label{alg:step3S}
		\State $(-1)^{f(i)}\ket{i}\ket{D}\ket{T_i}\ket{\rho(i)} \gets (-1)^{f(i)}\ket{i}\ket{D}\ket{T_i}\ket{0}$
		\State $(-1)^{f(i)}\ket{i}\ket{D}\ket{T_i}\ket{0}\gets (-1)^{f(i)}\ket{i}\ket{D}\ket{0}\ket{0}$
		\EndProcedure\label{alg:step3E}
		\Procedure{Diffusion Operator}{}\label{alg:step4S}
		\State $ \sum(-1)^{f(i)}\ket{i}  \gets \sum(2\ket{i}\bra{i}-\hat{\rm I})(-1)^{f(i)}\ket{i}$
		\EndProcedure\label{alg:step4E}
		\EndFunction
	\end{algorithmic}
\end{algorithm}

The Algorithm \ref{alg:GroGate} constructs a quantum oracle for Grover's algorithm, which plays the role of $U_f$ in  Eq.~\eqref{kick}. According to Ref.~\cite{gao2022quantum}, the quantum oracle begins with initialization, where the data is loaded and the index register is prepared in an equal superposition, at a cost of $O(M+\log N)$. The template waveforms are then generated from the index in superposition, costing $O(M)$, followed by the calculation of the SNR and comparison with a threshold, which requires $O(M\log M)$. To ensure the diffusion operator acts correctly on the index register, the intermediate computations of templates and SNR are erased, adding another $O(M\log M)$. Finally, the diffusion operator is applied, which requires $O(\log N)$. Hence, the total cost of a single oracle call is 
\begin{equation}
	\label{equ:oracletotalcost}
	O\left(M\log M + \log N\right).
\end{equation}

The Algorithm \ref{alg:GroPseudo} uses quantum counting to identify whether there is a signal existing in the data. Quantum counting returns $r_*$, an estimate of the number of matches, so it is suitable to do this task. According to Ref.~\cite{gao2022quantum}, there is a lower bound for the choice of number of counting qubits:
\begin{equation} \label{lower_bound}
	2^{p} > \pi\sqrt{N},
\end{equation}
and the signal detection algorithm based on quantum counting has a false alarm probability of 0 under all conditions, and a false negative probability of $1/\pi^2$, given the condition in Eq.~\eqref{lower_bound} is met. Here, the false-alarm and false-negative probabilities assume idealized arithmetic, perfect phase estimation, and flawless oracle construction. In realistic implementations, finite register precision and gate errors can shift the decision threshold, potentially increasing false-alarm and false-negative probabilities. While a detailed error-sensitivity analysis is beyond the scope of this work, we emphasize that the present study illustrates the algorithmic scaling and efficiency of quantum matched filtering under idealized conditions. The total cost of Algorithm \ref{alg:GroPseudo} is 
\begin{equation}
	\label{equ:al1totalcost}
	O\left((M\log M + \log N)\cdot\sqrt{N}\right).
\end{equation}

\begin{algorithm}[htbp]
	\caption{Signal Detection\newline Complexity: $O\left((M\log M + \log N)\cdot\sqrt{N}\right)$~\cite{gao2022quantum}}\label{alg:GroPseudo}
	\begin{algorithmic}[1]
		\State $\textit{p} \gets$ number of \textit{precision digits}
		\State $\textit{N} \gets$ number of \textit{templates}
		\State $i \gets $index of \textit{ templates}
		\State $\rho_{\textrm{thr}} \gets$ \textit{threshold}
		\State $\ket{0} \gets$ \textit{Data}  $\ket{D}$ 
		\Procedure{Quantum Counting}{$p$, $N$, $\ket{D}$, $\rho_{\textrm{thr}}$}
		\label{pro:QC}
		\State \emph{Creating the counting register }:\label{alg:2step1S}
		\State{$ \ket{i}\gets \ket{0}^p\ket{i}$}
		\State{$ \ket{0}^p\ket{i}\gets \frac{1}{2^{p/2}}(\ket{0}+\ket{1})^p\otimes\ket{i}$}
		\label{alg:2step1E}
		\State \emph{Controlled Grover's gate}:\label{alg:2step2S}
		\ForAll{$j<2^p$}
		\State $a \gets j$
		\Repeat
		\State Algorithm~\ref{alg:GroGate} \Call{Grover's Gate}{$N$, $\ket{D}$, $\rho_{\textrm{thr}}$}, $a--$
		\Until{$a==0$}
		\EndFor
		\State$\frac{1}{2^{p/2}}(\ket{0}+\ket{1})^n\otimes\ket{i} \gets \frac{1}{2^{(p+1)/2}}\sum ( e^{2i\theta j}\ket{j}\otimes\ket{s_+}+ e^{-2i\theta j}\ket{j}\otimes\ket{s_-})$\label{alg:2step2E}
		\State \emph{Inverse Quantum Fourier Transform}:\label{alg:2step3S}
		\State$\frac{1}{2^{(p+1)/2}}\sum ( e^{2i\theta j}\ket{j}\otimes\ket{s_+}+ e^{-2i\theta j}\ket{j}\otimes\ket{s_-}) \gets \frac{1}{2^{p+1/2}}\sum\sum( e^{i2\pi j(\frac{\theta}{\pi}-\frac{l}{2^p})}\ket{l}\otimes\ket{s_+}+ e^{i2\pi j(\frac{\pi-\theta}{\pi}-\frac{l}{2^p})}\ket{l}\otimes\ket{s_-})$\label{alg:2step3E}
		\State \emph{Measurement ($b$)}:\label{alg:2step4S}
		\If{$b=0$}
		\State \Return 'There is no match.'
		\Else{  $ r_\ast\gets \textbf{Round}\left[N\sin\left(\frac{b}{2^p}\pi\right)^2\right]$}
		\EndIf
		\If{$r_\ast=0$ }
		\State $r_\ast\gets 1$
		\EndIf\label{alg:2step4E} 
		\EndProcedure
	\end{algorithmic}
\end{algorithm}
Since the Algorithm \ref{alg:GroPseudo} returns an estimate of the number of matches, it is also easy to get $k_*$, an estimate of optimal number of Grover applications $k$ according to Eq.~\eqref{optk_appro}. After getting $k_*$, Algorithm \ref{alg:templateretreiving} uses Grover's algorithm to find a template. And the total cost of Algorithm \ref{alg:templateretreiving} is 
\begin{equation}
	\label{equ:al2totalcost}
	O\left((M\log M + \log N)\cdot\sqrt{N}\right).
\end{equation}

\begin{algorithm}[htbp]
	\caption{Template retrieval \newline Complexity: $O\left((M\log M + \log N)\cdot\sqrt{N}\right)$~\cite{gao2022quantum}}
	\label{alg:templateretreiving}
	\begin{algorithmic}[1]
		\State $\textit{N} \gets \textrm{number of }\textit{templates}$
		\State $i \gets \textrm{index of} \textit{ templates}$
		\State $\rho_{\textrm{thr}} \gets$ \textit{ threshold}
		\State $\ket{0} \gets \textit{Data}$  $ \ket{D}$ 
		\State $r_{\ast}\gets\textrm{number of }\textit{matched templates}$
		\State \emph{Calculating the number of repetitions}:\label{alg:3step0S}
		\State {$k_{\ast}\gets \textbf{Round}\left[\frac{\pi}{4}\sqrt{\frac{N}{r_{\ast}}}-\frac{1}{2}\right]$}\label{alg:3step0E}
		\Procedure{Retrieve one template}{}\label{alg:3step1S}
		\Repeat
		\State Algorithm~\ref{alg:GroGate} \Call{Grover's Gate}{$N$, $\ket{D}$, $\rho_{\textrm{thr}}$}, $k_{\ast}--$
		\Until{$k_{\ast}==0$}
		\State \emph{Output}:
		\State $i_{\textrm{correct}}$\label{alg:3step1E}
		\EndProcedure
	\end{algorithmic}
\end{algorithm}

Since scalable, error-corrected quantum processors are not yet available, it is necessary to summarize the space and gate requirements of the algorithm. According to Ref.~\cite{gao2022quantum}, for $N$ templates and $M$ time steps of signal data, the algorithm requires a counting register of size $\lceil \log_2 \pi + \frac{1}{2} \log_2 N \rceil$ qubits, an index register of $\log_2 N$ qubits, and two registers of $64 M$ qubits each (assuming 8 bytes per time sample) to hold the data and a template. As templates are stored in superposition, only one template register is needed. Generating templates and performing matched filtering reversibly add a modest logarithmic overhead in $M$. Overall, the data register dominates the space requirement. For the simulation example in \autoref{table_resource}, the time series spans 3 days and is sampled at 0.1 Hz, giving $M = 3 \times 24 \times 3600 / 0.1 = 2,592,000$. Assuming 8 bytes per data point, our algorithm is feasible on an error-corrected quantum device with approximately 20 MB of memory. For longer datasets spanning several months to years, the required memory increases linearly. For a year-long MBHB signal, approximately 2 GB of memory is required.

In the previous analysis, we assumed that the quantum FFT step inside the oracle has the same complexity as the classical FFT. A more realistic treatment would account for the additional overhead introduced by implementing the FFT quantumly. Any classical circuit with $T$ gates and $S$ bits can be converted into a reversible circuit with at most a polynomial overhead. Specifically, the number of gates can be upper bounded by $3T^{1+\Delta}$ for any $\Delta>0$~\cite{rieffel2011quantum}. Neglecting the small $T^\Delta$ factor and accounting for the need to erase intermediate calculations, the quantum oracle requires roughly 6 times as many gates as the corresponding classical FFT algorithm for computing the SNR.

We have considered only gate complexity. On nowdays error-corrected devices, quantum gates are slower and incur error-correction overhead, so quadratic speed-ups offer little runtime advantage. This is not a near-term application. However, future advances of the quantum hardware could enable quantum algorithms to enhance sensitivity in gravitational wave searches.

\begin{table}[h]
\centering
\caption{Estimated quantum resources for simulated case.}
\label{table_resource}
\begin{tabular}{lcc}
\hline
\textbf{Resource} & \textbf{Symbol} & \textbf{Estimate / Requirement} \\
\hline
Template number (library size) & $N$ & $2^{17}=131072$ \\
Data length & $M$ & 2592000 \\
Sampling rate & $f_s$ & $0.1$ Hz \\
Data register qubits & $n_d$ & $64 M \approx 20$MB \\
Counting register precision & $p$ & $11$ qubits \\
Index register qubits & $n_i$ & $\log_2 N = 17$ \\
Total qubits (logical) & $n_\text{tot}$ & $n_d + n_i + p \approx 20$ MB \\
Circuit depth (order) & $D$ & $O(\sqrt{N/r})$ Grover iterations \\
\hline
\end{tabular}

\end{table}

When simulating the quantum matched filtering algorithm for MBHB signals, there are challenges similar to those reported in the study of GW150914 in Ref.~\cite{gao2022quantum}. In particular, the large size of the input data and template bank exceeds the capacity of current quantum simulators, such as IBM’s Qiskit framework. Despite this, it remains feasible to compute the amplitudes of quantum states. A further complication lies in the construction of the quantum oracle, which must map each template index to a specific waveform. While an explicit quantum circuit for this task has not yet been implemented, theoretical results suggest that it can be constructed from its classical counterpart without increasing the asymptotic complexity of the overall algorithm~\cite{bennett1997strengths}. Following the method in Ref.~\cite{gao2022quantum}, the implementation circumvents this difficulty by employing a precomputed lookup table: each input index is mapped to a corresponding parameter pair ${m_1}$, ${m_2}$ that defines the waveform. In a quantum implementation, this step is not performed via a lookup table, as that would require qRAM. The simulation is built upon the \href{https://github.com/Fergus-Hayes/quantum-matched-filter}{\textit{quantum-matched-filter}} \textit{Python} code~\cite{pythoncode}, which accompanies Ref.~\cite{gao2022quantum}. The gravitational wave strain data used in this study is generated using BBHX package~\cite{katz2020gpu, katz2022fully, michael_katz_2021_5730688} and simulates the signal from a binary black hole system with component masses of $10^6\,M_\odot$ and $5 \times 10^5\,M_\odot$. The time series spans 3 days and is sampled at 0.1~Hz.

\autoref{table_resource} lists some of the requirements and estimated quantum resources for the simulated case.
 The analysis is performed on a bank of $2^{17}$ templates covering the 2-dimensional search space defined by the masses $m_1$, $m_2$ of the binary system. To search these templates to find instances that correspond to matching templates, first the \textsc{Signal Detection} procedure of Alg.~\ref{alg:GroPseudo} is applied to determine if a signal is present in the data, and obtain an estimation on the number of matching templates.

\begin{figure} [htbp]
	\centering
	\includegraphics[scale=0.32]{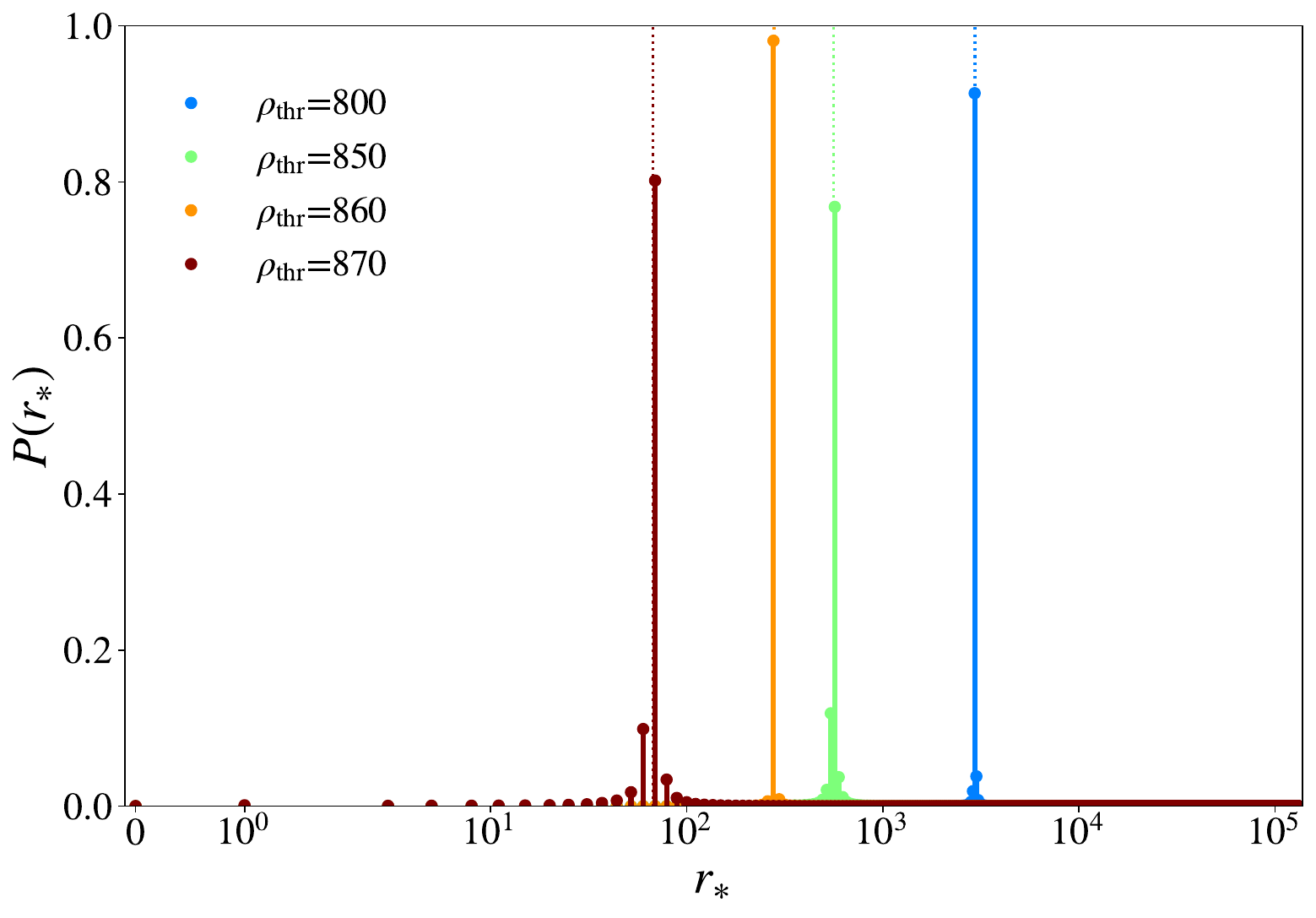}
	\caption{The probability distributions of outcomes from measuring the counting register transformed to estimates on the 
 number of matching templates $r_\ast$ for each of the different cases of $\rho_{\text{thr}}$. The distributions are compared to the true number of matching templates $r$ (dotted).}
	\label{r} 
\end{figure}
\begin{figure} [htbp]
	\centering
	\includegraphics[scale=0.32]{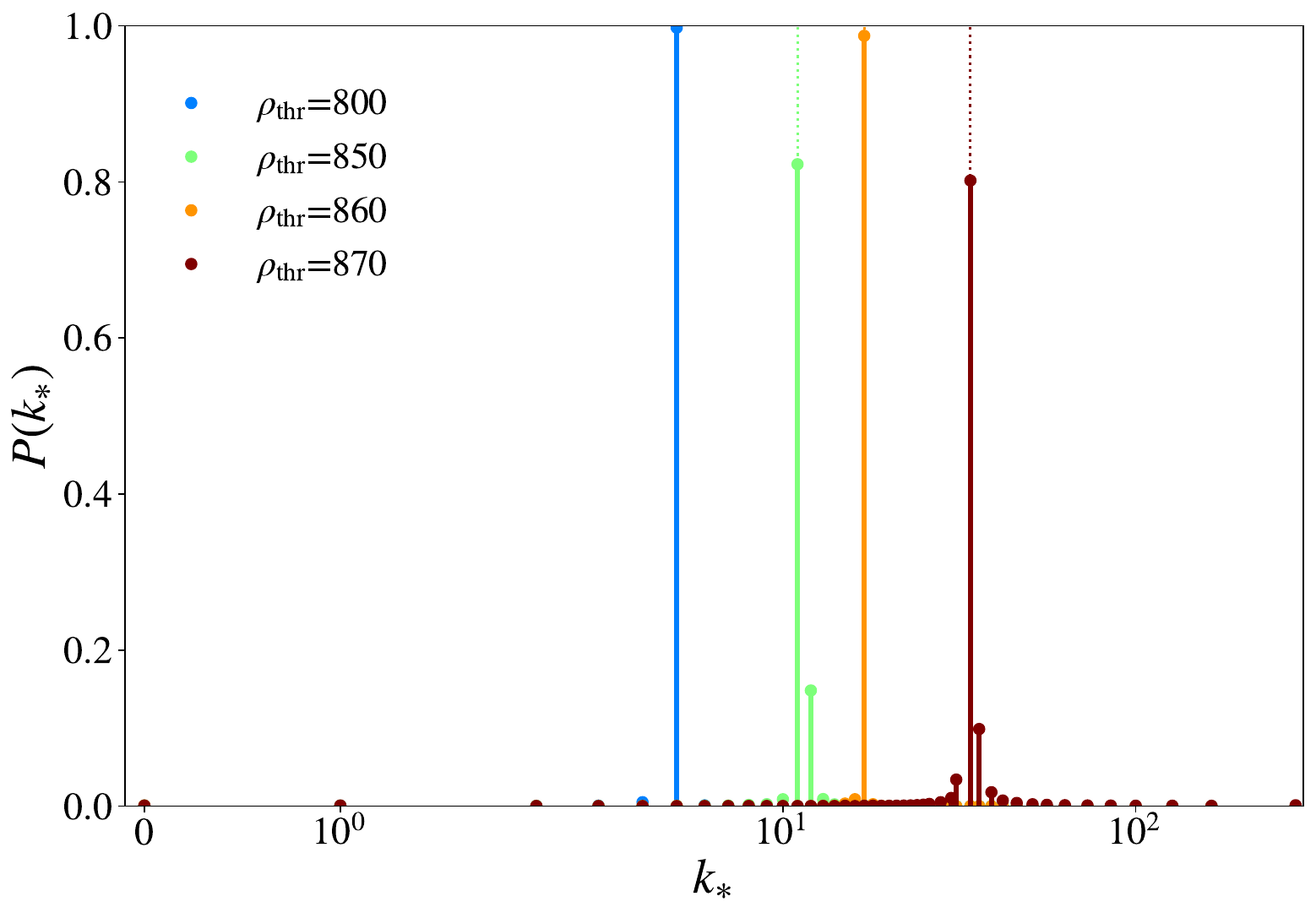}
	\caption{The probability distributions of outcomes from measuring the counting register transformed to estimates on the optimal number of Grover’s applications $k_\ast$ for each of the different cases of $\rho_{\text{thr}}$. The probabilities are compared to the true $k$ (dotted) for each case.}
	\label{k} 
\end{figure}
\begin{figure*}
\centering
\subfloat[$\rho_{\text{thr}}=800$]{\label{figure_sima}\includegraphics[height=0.3\textwidth]{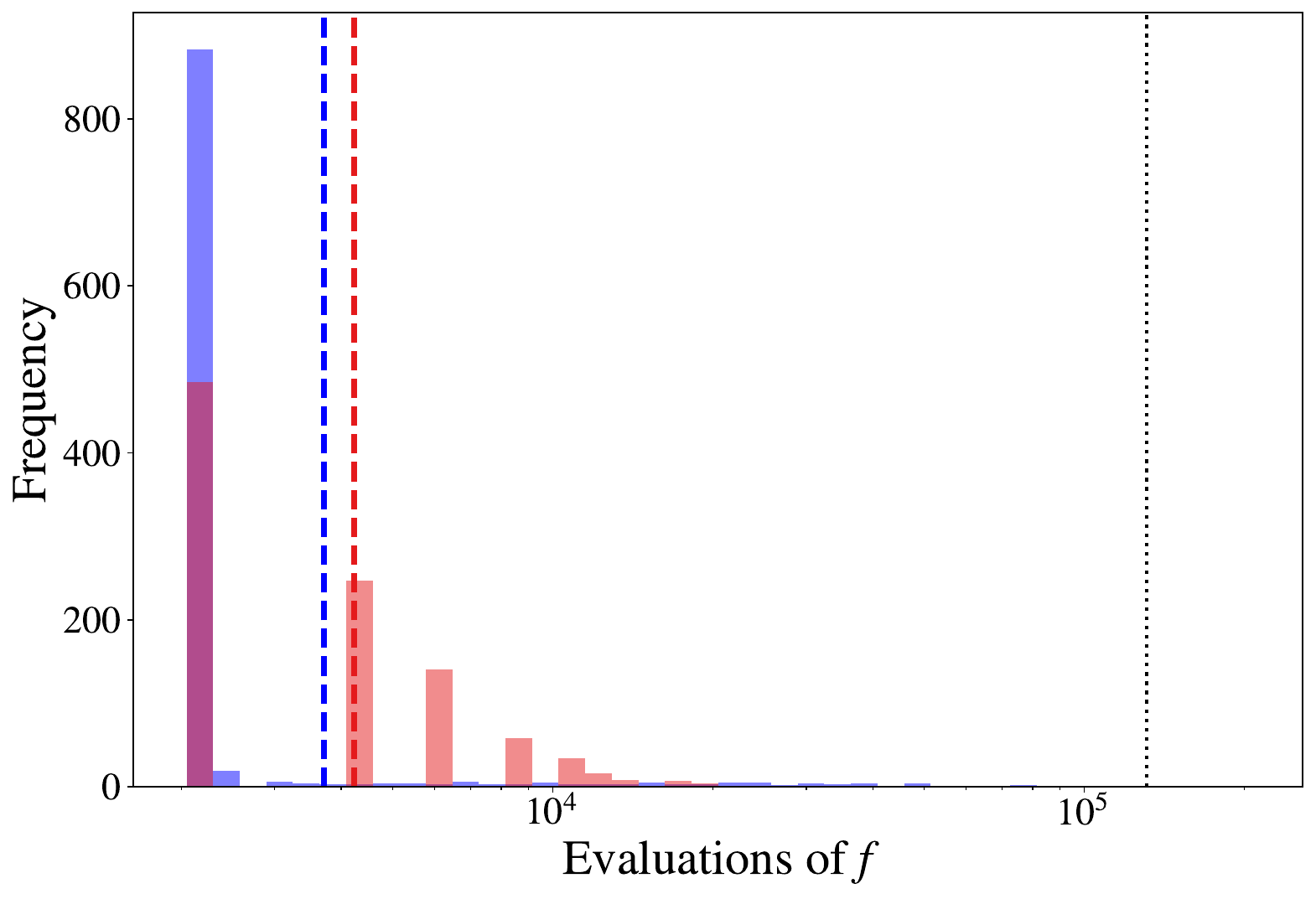}}
\subfloat[$\rho_{\text{thr}}=850$]{\label{figure_simb}\includegraphics[height=0.3\textwidth]{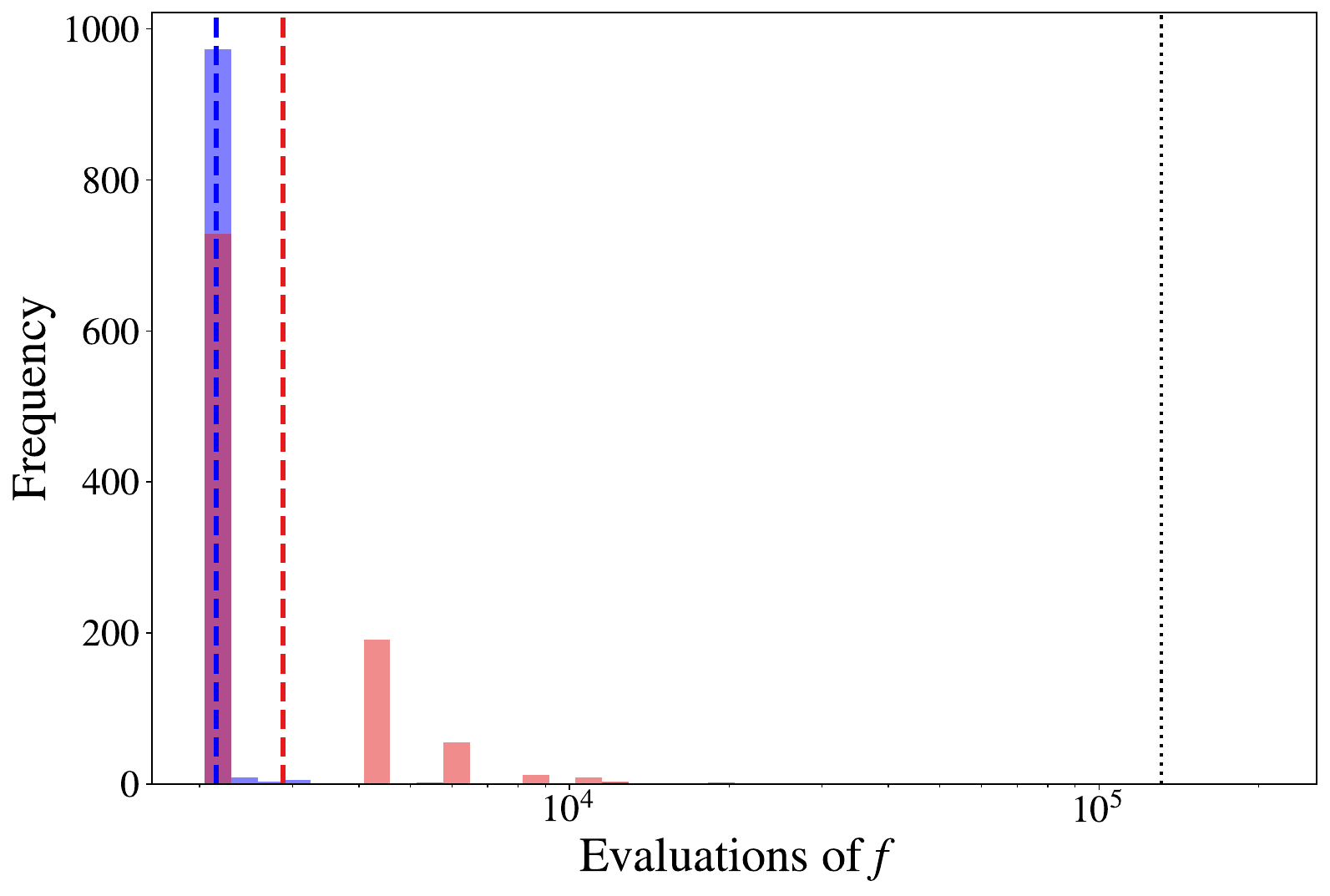}}\\
\subfloat[$\rho_{\text{thr}}=860$]{\label{figure_simc}\includegraphics[height=0.3\textwidth]{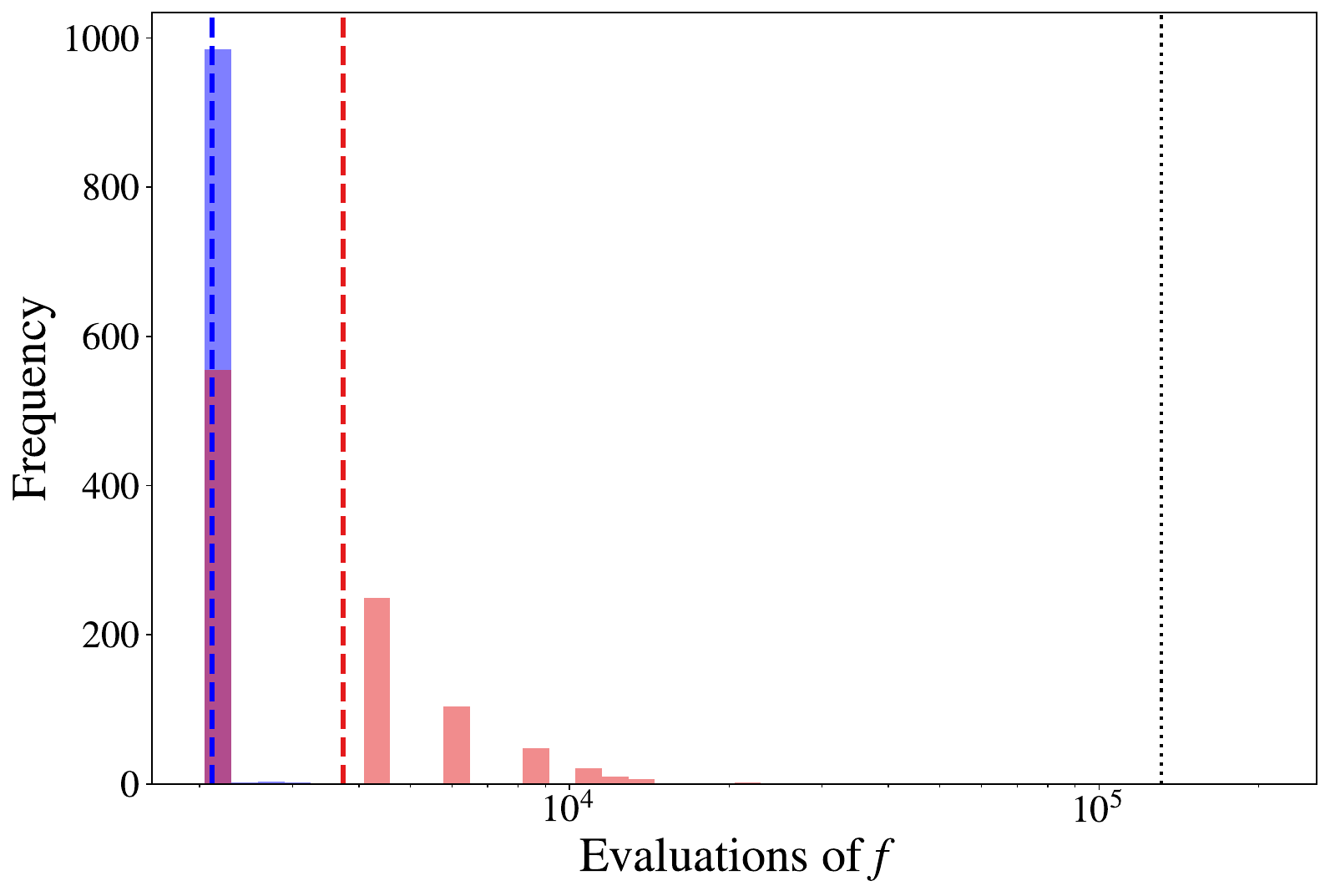}}
\subfloat[$\rho_{\text{thr}}=870$]{\label{figure_simd}\includegraphics[height=0.3\textwidth]{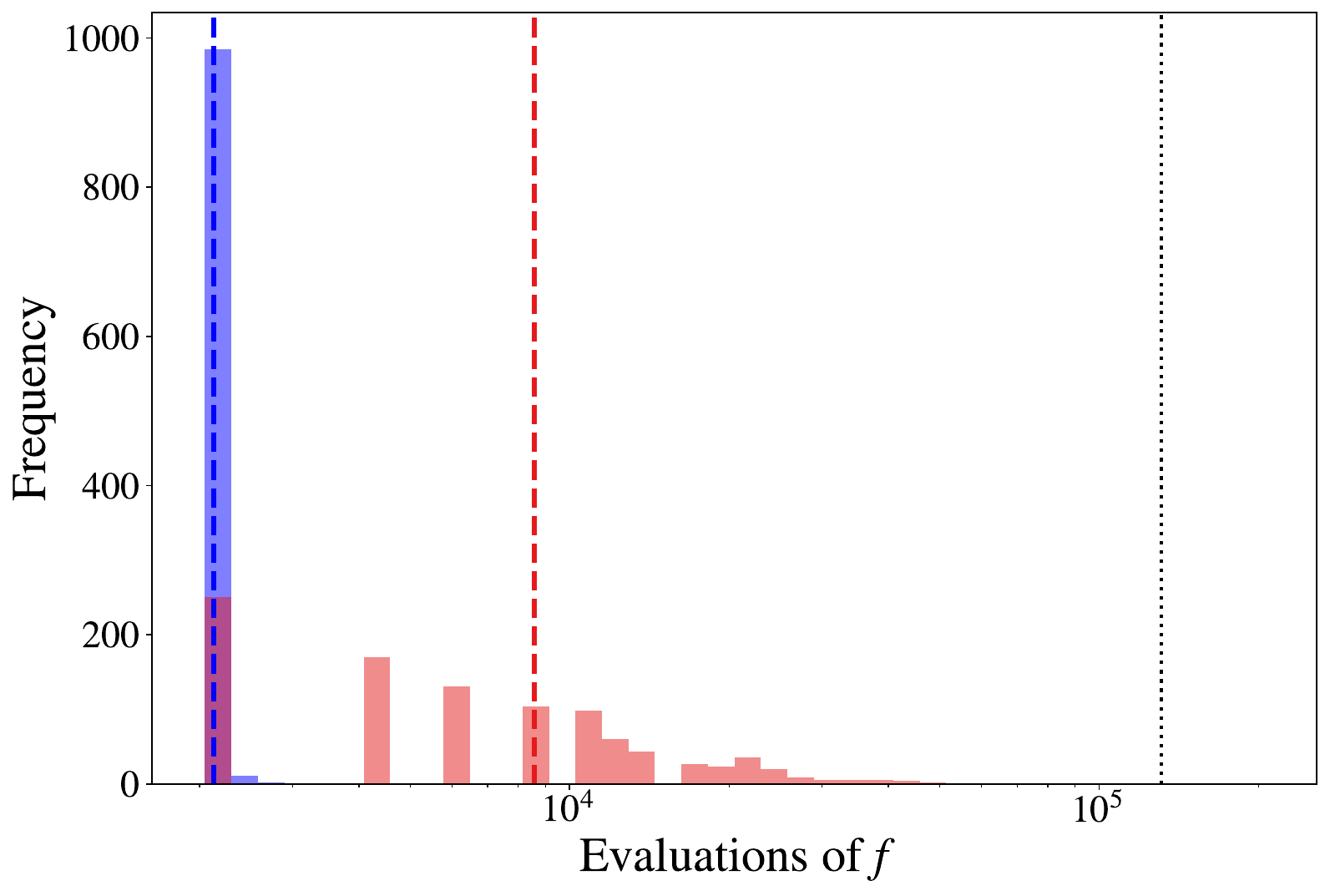}}
\caption{Number of evaluations of the oracle function $ f $ required to retrieve a matching template in 1,000 simulations at different threshold levels \( \rho_{\text{thr}} = 800, 850, 860, 870 \). (a–d) Each subplot shows the distribution of function evaluations across trials. Blue histograms represent using a fixed $ k_* $ estimated from a single \textsc{Signal Detection} step; red histograms correspond to re-estimating $ k_* $ for each failed retrieval. Dashed lines indicate mean values. The black dotted line shows the classical case where all $2^{17}$ templates are evaluated.
}
\label{fig:sims}
\end{figure*}
Since the template bank size is $2^{17}$, the number of qubits in the counting register is set to $p=11$, which is appropriate for Eq.~\eqref{lower_bound} and provides a good level of precision. Consequently, the oracle is invoked a total of $(2^{11} - 1)$ times under the control of the counting register. After applying the inverse quantum Fourier transform to the counting register, an estimate of $\theta$ can be obtained, which is then used to calculate the estimated number of matching templates $r_*$. In the simulation, four threshold values are selected, $\rho_{\text{thr}} = 800, 850, 860, 870$, corresponding to $r$ values covering the range from $10$ to $10^3$. As shown in Fig.~\ref{r}, the probability distributions of the estimated values $r_*$ for these four cases are presented. The dashed vertical lines represent the true number of matching templates $r$, as determined by classical matched filtering. It can be seen that the estimated value $r_*$ corresponding to the maximum probability is very close to the true value $r$, while the probabilities of other $r_*$ values are relatively low. Therefore, a measurement is likely to yield an estimate with a small error, indicating the efficiency of signal detection.

Since $r$ is known by quantum counting, the optimal number of Grover’s operations is estimated using Eq.~\eqref{optk_appro}. Figure~\ref{k} shows the probability distributions of the estimated values $k_*$ for the four cases mentioned before. The dashed vertical lines represent the true number of matching templates $k$.

After obtaining $k_{\ast}$, the \textsc{Template Retrieval} procedure in Alg.~\ref{alg:templateretreiving} is applied to find a matching template. If the procedure fails, it is repeated using the same $k_{\ast}$ until a matching template is found. We conduct 1000 simulations for this case in which $k_{\ast}$ is estimated via the \textsc{Signal Detection} procedure, followed by repeated applications of \textsc{Template Retrieval} until a match is identified. The number of function evaluations in each simulation is shown in the blue histogram of Fig.~\ref{fig:sims}, with the average indicated by the blue dashed line. For comparison, the classical exhaustive search is shown as the black dotted line.

If repeated retrieval attempts fail, one may assume that the estimated $k_{\ast}$ is suboptimal and rerun the \textsc{Signal Detection} procedure to obtain a new estimate. The number of function evaluations under this alternative approach—where a new $k_{\ast}$ is generated for each failed attempt—is shown in the red histogram of Fig.~\ref{fig:sims}, which exhibits a higher cost compared to the original method. Despite the increased cost, the mean number of function evaluations in this case still remains lower than that of a full classical search. Although the quantum algorithm requires fewer oracle calls, this does not necessarily imply that its total computational resources are lower, since the cost of each oracle call can be significant. A more accurate comparison of their computational costs would need to take into account the resources consumed by a single oracle call. Since the construction of the quantum oracle is difficult to simulate, a precise evaluation of the oracle cost is beyond the scope of our simulations. Nevertheless, previous theoretical analyses suggest that the resource cost of a quantum oracle is expected to scale polynomially higher than that of its classical counterpart. Despite this potential overhead, with anticipated progress in quantum hardware, the $\sqrt{N}$ speedup of Grover’s algorithm may still hold potential for practical applications.

Considering the average performance of the algorithm, quantum approach require fewer oracle calls than its classical counterparts. However, quantum algorithms still have room for improvement. Specifically, in the cases of $\rho_{\text{thr}} = 800$ and $870$, the run cost distributions exhibit long tails, indicating that a small fraction of simulations take significantly longer than the average. The reasons behind this behavior are not yet fully understood, suggesting that the robustness of current quantum algorithms needs further enhancement.
A closer examination reveals that these inefficiencies manifest differently in the two scenarios. In the case of $\rho_{\text{thr}} = 800$, the blue cost distribution shows a pronounced long tail. This is further evidenced by the fact that its mean is significantly shifted to the right of the peak, in contrast to typical distributions where the mean is located near the peak.
In the case of $\rho_{\text{thr}} = 870$, the red distribution also displays a significant long tail, resulting in a rightward shift of its mean. Notably, the gap between the mean of the red distribution and that of the blue distribution is substantially larger than in typical cases, further highlighting the increased variability and instability under this condition.

\begin{figure}
\centering
\subfloat[$\rho_{\text{thr}}=18$]{\label{original_result}\includegraphics[height=0.3\textwidth]{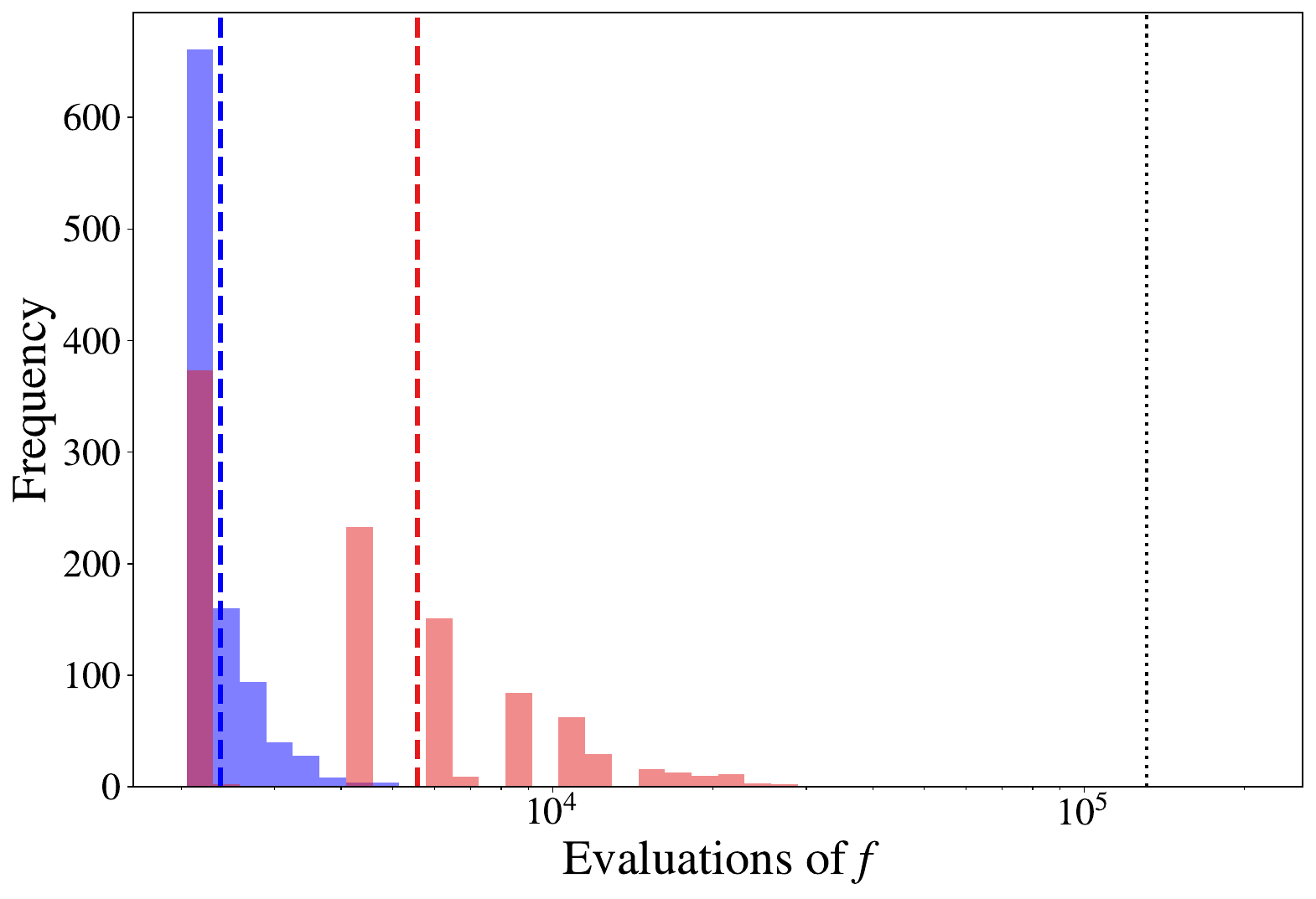}}\\
\subfloat[$\rho_{\text{thr}}=16$]{\label{thr_16}\includegraphics[height=0.3\textwidth]{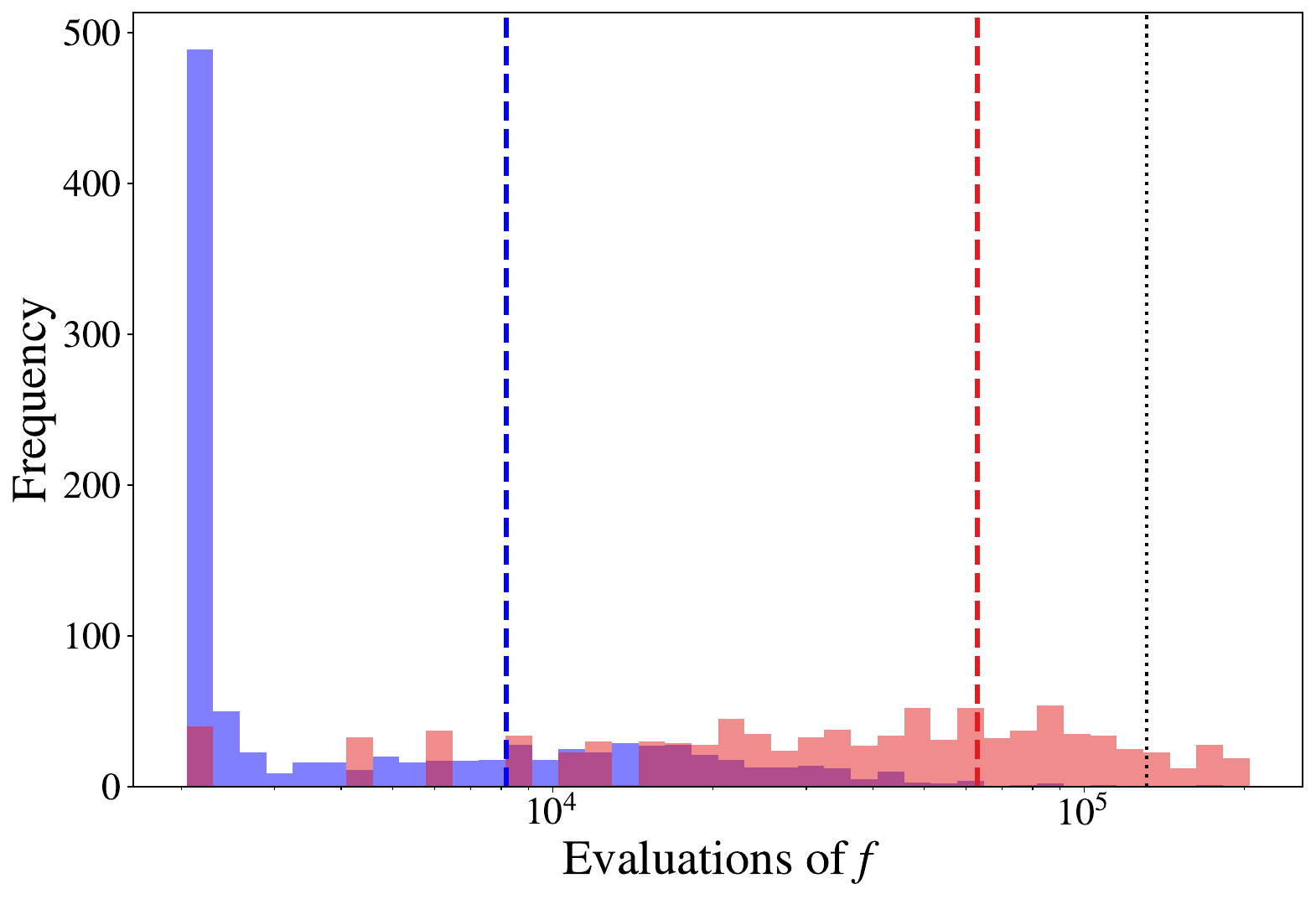}}
\caption{Comparison of algorithm efficiency for detection thresholds $\rho_{\text{thr}} = 16$ and $\rho_{\text{thr}} = 18$ using 1000 simulations given the GW150914 example. The $\rho_{\text{thr}} = 18$ case reproduces the results reported in Ref.~\cite{gao2022quantum}, while the $\rho_{\text{thr}} = 16$ case shows the results when lowering the threshold. This comparison highlights the sensitivity of the quantum matched filtering algorithm’s performance to the choice of threshold.}
\label{reproduce}
\end{figure}

A similar phenomenon is observed when reproducing the results on GW150914 from Ref.~\cite{gao2022quantum}. Using exactly the same parameters described in Ref.~\cite{gao2022quantum}, the results is shown in Fig.~\ref{original_result}, consistent with the original report of Ref.~\cite{gao2022quantum}. However, a slight change in the threshold, from $\rho_{\text{thr}} = 18$ used in Ref.~\cite{gao2022quantum} to $\rho_{\text{thr}} = 16$, leads to a significant drop in the algorithm's efficiency, which is shown in Fig.~\ref{thr_16}. In both cases, the quantum algorithm does not appear to yield satisfactory performance. Although the quantum algorithm remains more efficient on average, the long tail observed in the red distribution indicates that, in certain instances, its performance can even fall below that of the classical approach. This again emphasizes the sensitivity of current quantum algorithms to $\rho_{\text{thr}}$, and underscores the importance of further improving their stability and robustness.  

Here is an incomplete possible explanation: the threshold determines the number of marked solutions $r$ within a search space of size $N$. As $\rho_{\text{thr}}$ increases, fewer candidate templates exceed the threshold and $r$ decreases; conversely, lowering $\rho_{\text{thr}}$ increases $r$.

A smaller $r$ implies more iterations and thus higher computational cost. On the other hand, a larger $r$ reduces the required iterations but may cause amplitude amplification to deviate from the ideal rotation, lowering the success probability per measurement. This can lead to the need for more repeated runs to ensure reliable detection, potentially increasing overall computational cost.

The results on MBHB signals show that the efficiency is higher at intermediate thresholds ($\rho_{\text{thr}} = 850, 860$) and decreases at the extreme thresholds ($\rho_{\text{thr}} = 800, 870$). This is consistent with the analysis presented above. However, the situation for GW150914 appears to be more complex. 

\begin{figure*}
\centering
\subfloat[$p=10$]{\label{figure_sima}\includegraphics[height=0.2\textwidth]{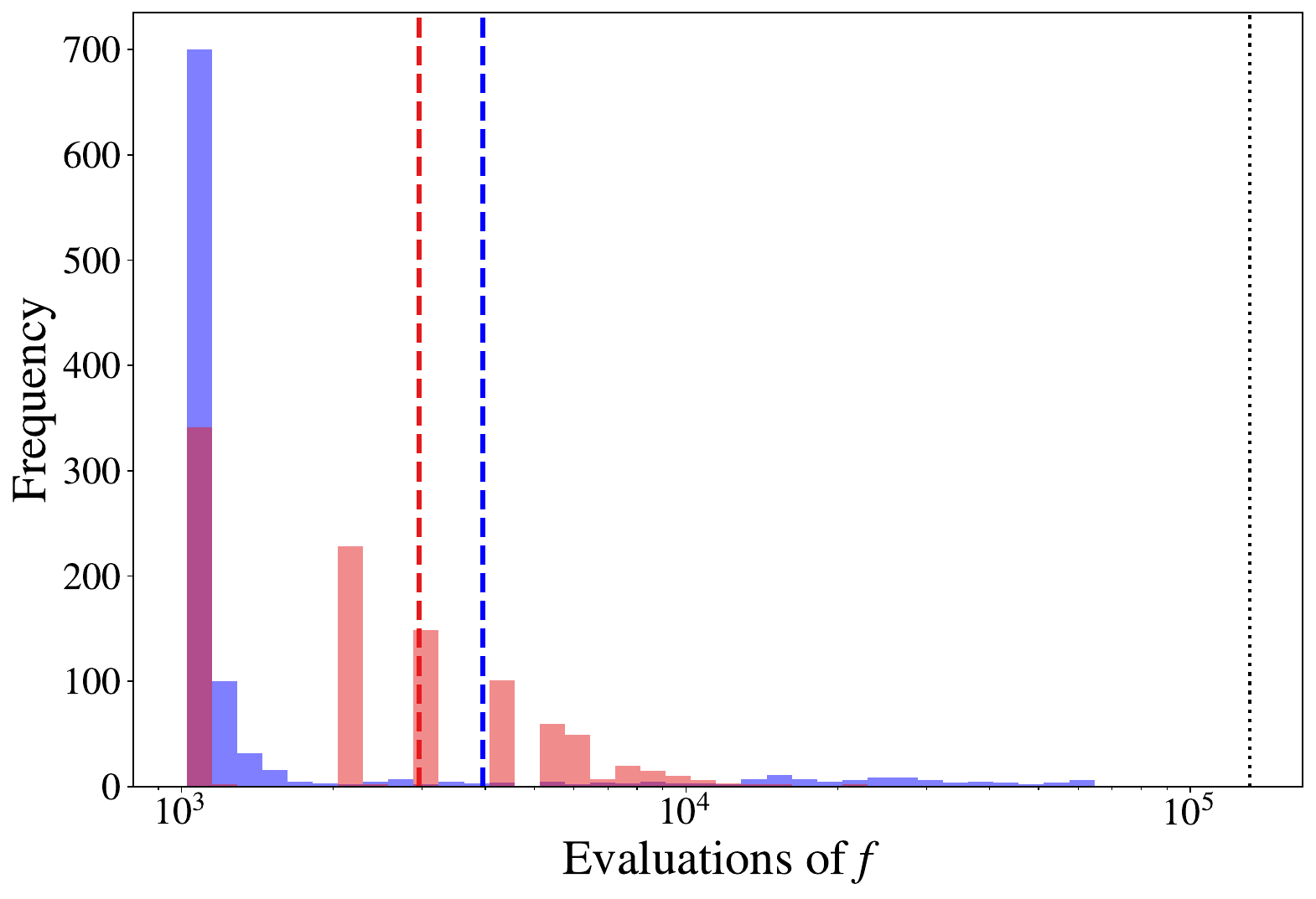}}
\subfloat[$p=11$]{\label{figure_simb}\includegraphics[height=0.2\textwidth]{131072_2048_800_simulation_scenarios.pdf}}
\subfloat[$p=12$]{\label{figure_simc}\includegraphics[height=0.2\textwidth]{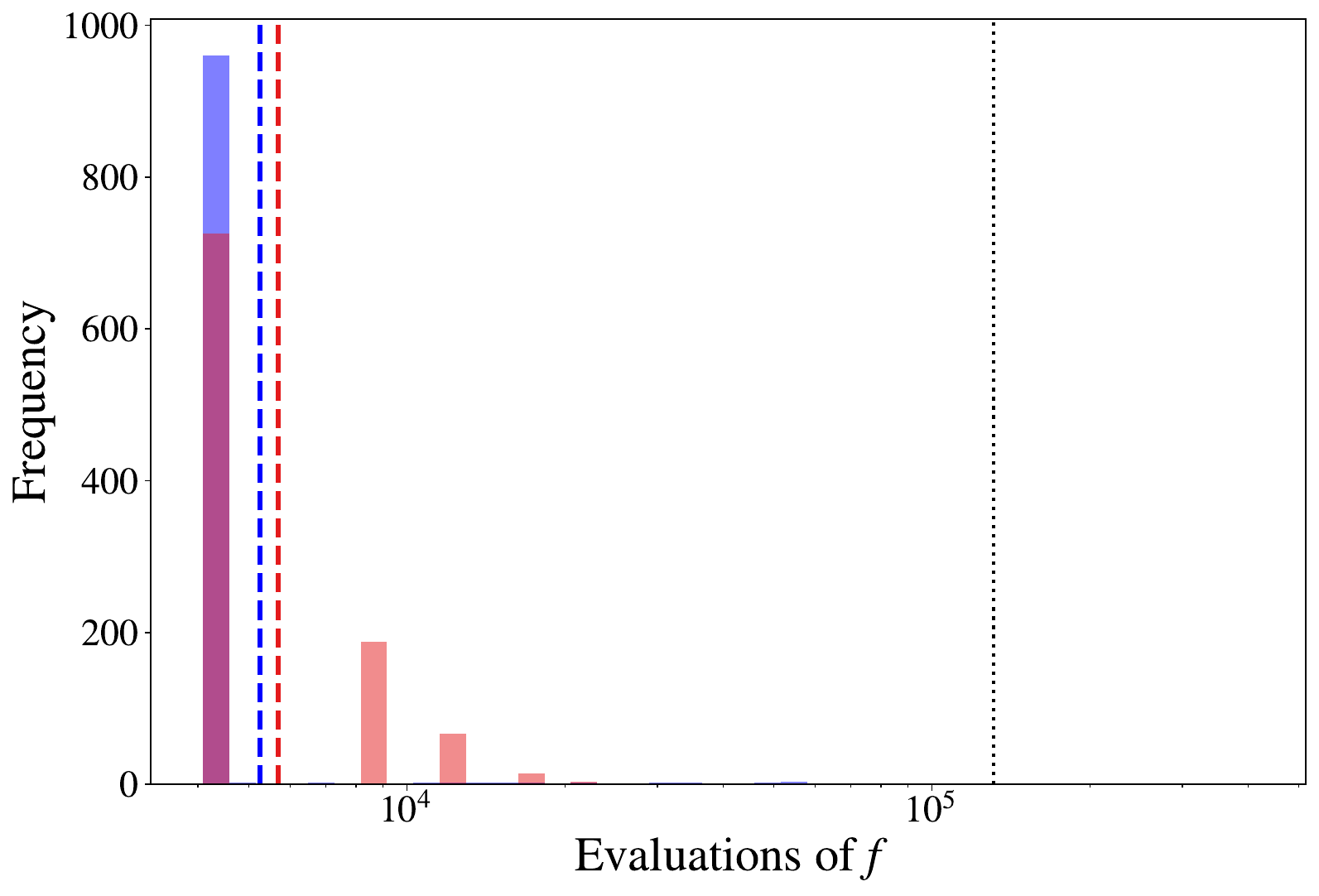}}
\caption{Number of oracle function $f$ evaluations required to retrieve a matching template in 1,000 simulations, for precision values $p = 10, 11, 12$, with the threshold set to $\rho_\text{thr}=800$.
}
\label{fig:sims800_p_vary}
\end{figure*}

\begin{figure*}
\centering
\subfloat[$p=10$]{\label{figure_sima}\includegraphics[height=0.2\textwidth]{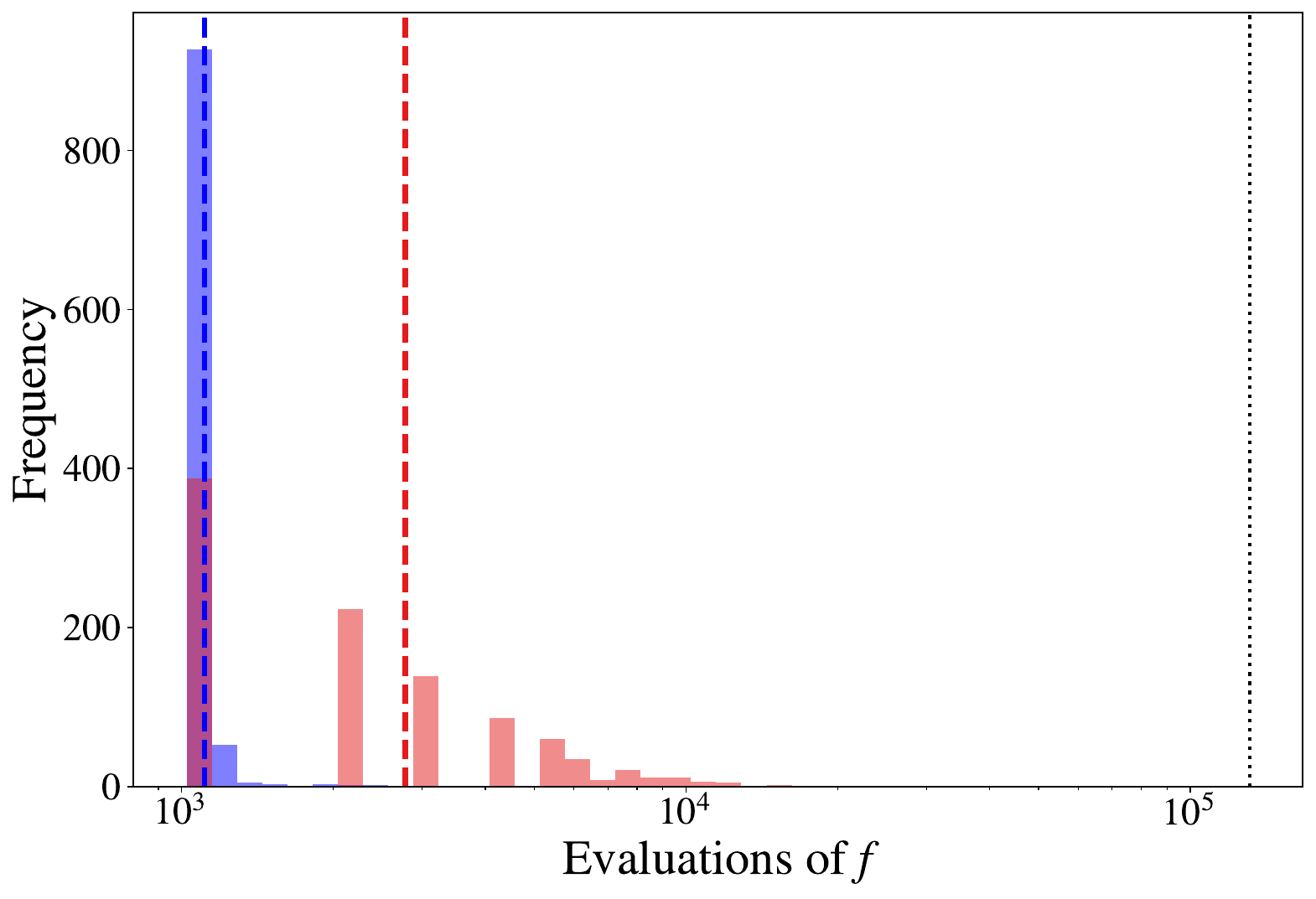}}
\subfloat[$p=11$]{\label{figure_simb}\includegraphics[height=0.2\textwidth]{131072_2048_870_simulation_scenarios.pdf}}
\subfloat[$p=12$]{\label{figure_simc}\includegraphics[height=0.2\textwidth]{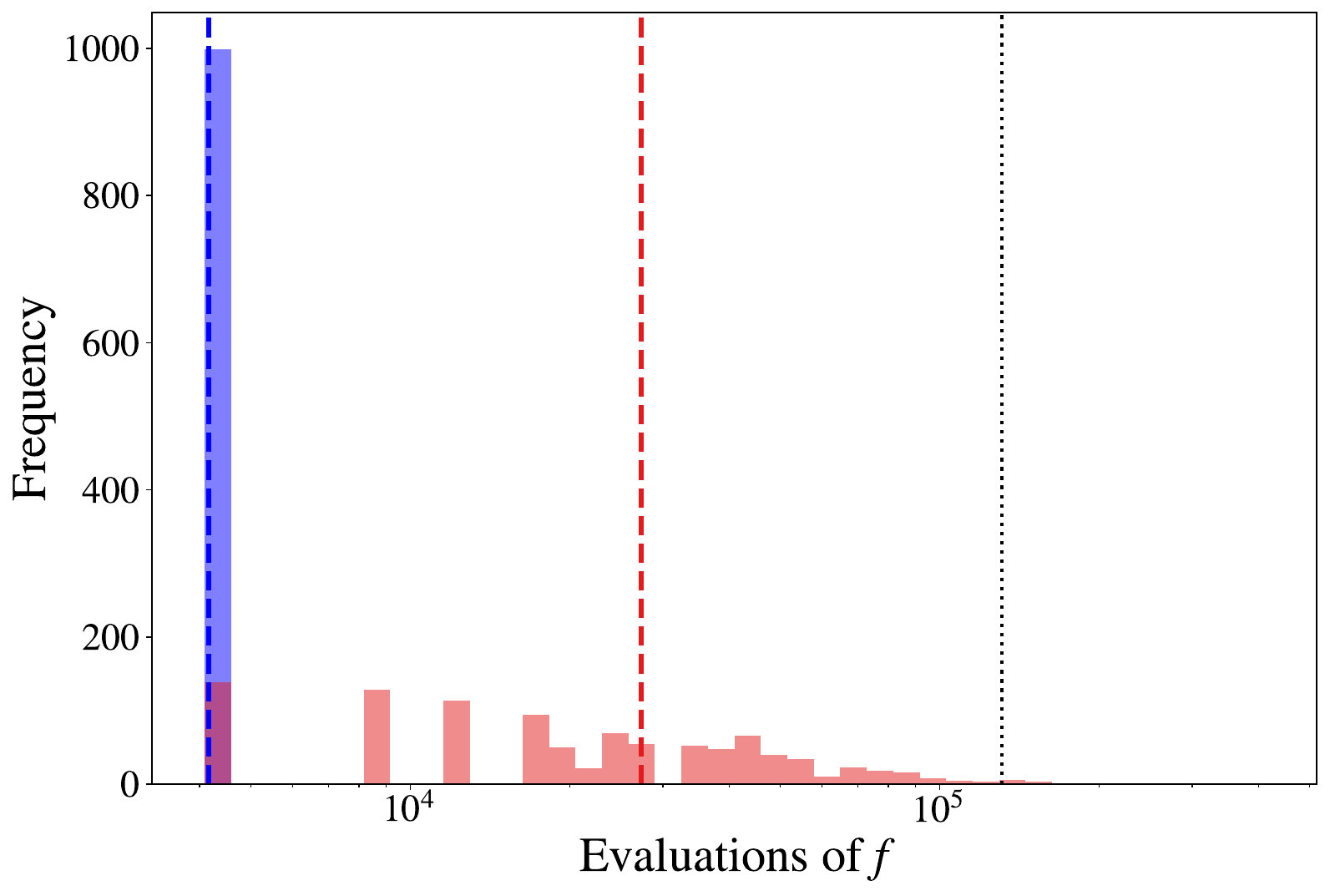}}
\caption{Number of oracle function $f$ evaluations required to retrieve a matching template in 1,000 simulations, for precision values $p = 10, 11, 12$, with the threshold set to $\rho_\text{thr}=870$.
}
\label{fig:sims870_p_vary}
\end{figure*}

In the simulations shown in Fig.~\ref{fig:sims}, the precision $p$ was set to 11, which corresponds to the minimum value satisfying Eq.~\ref{lower_bound}. To examine the impact of different quantum counting precision values on the algorithm's performance, and to see whether increasing or decreasing $p$ could improve the relatively poor performance observed for the thresholds $\rho_\text{thr} = 800$ and $870$, we conducted additional tests for these two thresholds with $p = 10, 11, 12$. The results are shown in Figs.~\ref{fig:sims800_p_vary} and \ref{fig:sims870_p_vary}. As can be seen, for $\rho_\text{thr}=800$ increasing the precision to $p=12$, and for $\rho_\text{thr}=870$ decreasing it to $p=10$, both lead to improved performance compared with the original settings. 

In Figs.~\ref{fig:sims800_p_vary} and \ref{fig:sims870_p_vary}, it can be observed that as the precision $p$ increases, the long-tail behavior of the blue bars decreases, while the peak of the distribution becomes higher. This indicates that the distribution becomes more concentrated, and hence the algorithm's robustness improves. Theoretically, increasing $p$ leads to higher precision in quantum counting, which in turn provides a more accurate estimate of $k$, resulting in a higher success probability for Grover's algorithm. Consequently, the algorithm is more likely to find the matching template in a single attempt, meaning that its robustness increases with $p$. This observation agrees well with theoretical expectations. 

For Fig.~\ref{fig:sims800_p_vary}, the long-tail behavior of the red bars decreases as $p$ increases, whereas in Fig.~\ref{fig:sims870_p_vary} the long-tail effect of the red bars becomes more pronounced with larger $p$. We interpret this as follows. Increasing $p$ enhances robustness but also incurs higher computational cost. For the case of $\rho_\text{thr}=800$, when $p=10$ the robustness is still insufficient; thus, increasing $p$ mainly contributes to improving robustness, and at $p=12$ the overall performance becomes satisfactory. In contrast, for $\rho_\text{thr}=870$, the robustness is already adequate at $p=10$, and the overall performance is acceptable. Further increasing $p$ in this case mainly increases the overhead, which leads to degraded performance. 

The current testing scenarios remain limited. To gain a better understanding of how the value of $r$, $p$ and other factors influence the algorithm’s efficiency, further experiments are needed, for example, by scanning $r$ values with fixed step sizes to systematically examine how the algorithm performance varies with $r$. The extent to which the value of $r$ affects the algorithm’s performance, as well as a quantitative characterization of this relationship, requires further investigation. Additionally, other potential factors influencing efficiency also need deeper study. Besides, the simulation results and analysis indicate that our current theoretical treatment of the choice of $p$ in Eq.~\ref{lower_bound} remains incomplete, and a more refined theoretical analysis will be needed in future work. Moreover, in practical implementations, determining how to select an appropriate $p$ for different thresholds so as to achieve a suitable balance between robustness and performance is an important question that deserves further investigation. Another potential direction for improving robustness is to consider using variants of Grover’s algorithm that are less sensitive to the precise value of $r$ such as the Grover-Long algorithm~\cite{long2001grover}. It may offer more stable performance. Unlike the standard Grover algorithm, which suffers from over-rotation and reduced success probability when $r_\ast$ is inaccurate or too large, the Grover-Long algorithm theoretically achieves 100\% convergence to the solution, though with increased query complexity.

\section{Discussion} \label{sec:4}
In this work, the quantum matched filtering algorithm proposed by Ref.~\cite{gao2022quantum} is applied to MBHB signals. We have considered a simplified search simulation to illustrate the algorithmic framework of quantum matched filtering. A full resource analysis in higher-dimensional MBHB parameter spaces—accounting for distance, inclination, sky localization, spin, and other astrophysical parameters—remains beyond the present scope. We emphasize that our results should therefore be interpreted as a proof-of-principle demonstration, rather than a claim of immediate feasibility in realistic gravitational-wave searches. Extending the framework to more realistic waveforms and parameter spaces represents an important direction for future research. Moreover, improving the robustness of quantum algorithms remains an open challenge. Besides, the quantum algorithm is still far from practical implementation. In the algorithm, the core component—the construction of the oracle, which includes the matched filtering process, the generation of templates from template indices, and other steps—is simulated using classical methods, without a corresponding quantum circuit implementation. Moreover, aspects such as template waveform encoding and the parallel loading of template waveforms require corresponding improvements in quantum circuit design. These are some challenging aspects of implementing this algorithm and remain to be further explored. 

Besides, a possible application of this algorithm is the search for Extreme Mass Ratio Inspirals (EMRIs), which are among the key sources targeted by LISA and Taiji.
Compared to MBHBs, EMRI signals pose even greater challenges for data analysis. EMRI systems have a high-dimensional parameter space, often involving more than a dozen intrinsic and extrinsic parameters. This leads to an exponential growth in the number of templates required for matched filtering. It is estimated that template-based search for EMRI signals, with orders of magnitude more cycles in band, would require about $10^{40}$ templates to cover the parameter space, much bigger than MBHBs, rendering the approach unfeasible~\cite{gair2004event}.

The challenges make EMRIs a difficult target for traditional matched filtering methods. The quantum matched filtering algorithm, which offers parallel search capabilities and potential computational speedups in high-dimensional spaces, may help address some of these limitations. By encoding a large template space into quantum states and applying amplitude amplification techniques, quantum matched filtering may improve search efficiency.

\begin{acknowledgments}
This research is supported by 
the Research Funds of Hangzhou Institute for Advanced Study, UCAS, and partly 
funded by the Strategic Priority Research Program of the Chinese Academy of Sciences under Grant No. XDA15021100, and the Fundamental Research Funds for the Central Universities.

\end{acknowledgments}

\bibliography{ref_revise}

\end{document}